\documentclass[aps,prd,preprint]{revtex4}
\usepackage{amsmath}
\usepackage{relsize}

\newcommand{\beqa}{\begin{eqnarray}}
\newcommand{\eeqa}{\end{eqnarray}}
\newcommand{\beq}{\begin{equation}}
\newcommand{\eeq}{\end{equation}}
\newcommand{\bsp}{\begin{split}}
\newcommand{\esp}{\end{split}}
\newcommand{\bal}{\begin{align}}
\newcommand{\eal}{\end{align}}

\begin{document}

\title{Towers of positive parity excited baryons and their mixing in the large $N_c$ limit}

\def\addCW{Departamento de F\'{\i}sica, FCEyN, Universidad de Buenos Aires and IFIBA, 
\\ CONICET, Ciudad Universitaria, Pabell\'on~1, (1428) Buenos Aires, Argentina}
\def\addCS{\"Okumenisches Gymnasium zu Bremen, \\  Oberneulander Landstr.143a,  28355 Bremen, Germany}

\author{Cintia Willemyns}
\email{cintiawillemyns@gmail.com}
\affiliation{\addCW}
\author{Carlos Schat}
\email{carlos.schat@gmail.com}
\affiliation{\addCS}  

\begin{abstract} \vspace*{18pt}

We consider configuration mixing for the nonstrange positive parity excited
baryons in the $[\mathbf{56'},0^+],[\mathbf{56},2^+], [\mathbf{70},0^+] $ and
$[\mathbf{70},2^+]$ quark model $SU(6) \times O(3)$ multiplets contained in the
$N=2$ band. Starting from the effective mass operator for these states we show
by an explicit calculation  that in the large $N_c$ limit they fall into six
towers of degenerate states labeled by $K=0,1,1',2,2',3$. We find that the
mixing of the quark model states is much simpler than what is naively expected.
To leading order in $N_c$ only states carrying the same $K$ label can mix,
which implies that for the spin-flavor states we started with configuration
mixing can be parameterized by just two constants, $\mu_{K=1}$ and $\mu_{K=2}$.

\end{abstract} 

\date{\today}

\maketitle

\section{Introduction}

The quark model in its different versions has been a useful tool in exploring
the spectrum and properties of excited baryons, as well as for testing
different model assumptions on the effective interactions between quarks
\cite{Capstick:2000qj,Klempt:2009pi,Crede:2013kia}.  Although the model
interactions between constituent quarks are QCD inspired, a clear connection to
the fundamental theory of the strong interactions remains elusive to this date.
Recent successes of lattice QCD calculations
\cite{Edwards:2011jj,Edwards:2012fx} seem to validate the classification scheme
of baryon states in $SU(6)\times O(3)$ multiplets, as put forward by early
quark model studies.  These numerical lattice calculations have the advantage
of being based on the fundamental theory of the strong interactions, but lack
the simplicity of an analytic approach, that provides a physical picture in
terms of effective degrees of freedom, effective interactions and symmetries.
Such an analytic scheme to study the phenomenology of baryons and their excited
states, that also makes contact with QCD, can be obtained starting from the large
number of colors ($N_c$) limit of QCD \cite{'tHooft:1973jz}
\cite{Witten:1979kh}. 
For a review on the relevance of the large $N_c$ limit for $SU(N_c)$ 
gauge theories, see also Ref.~\cite{Lucini:2012gg}.

In the large $N_c$ limit it has been shown that the
spin-flavor symmetry for ground state baryons can be justified from the
contracted symmetry $SU(4)_c$ derived from consistency relations for
pion-nucleon scattering  \cite{Gervais:1983wq} \cite{Gervais:1984rc}
\cite{Bardakci:1983ev} \cite{Dashen:1993as} \cite{Jenkins:1993zu}.
The predictions of this symmetry for the masses and the couplings explain some
of the successes of the non-relativistic quark model \cite{Dashen:1993jt}
\cite{Carone:1993dz} \cite{Luty:1993fu} 
\cite{Dashen:1994qi}. Even more important than that, the breaking of the
spin-flavor symmetry can be studied systematically in a $1/N_c$ expansion
using quark operators, establishing a close connection between QCD and the
quark model \cite{Carone:1993dz} \cite{Luty:1993fu} \cite{Dashen:1994qi}, 
see Refs.~\cite{Manohar:1998xv,Jenkins:1998wy,Lebed:1998st} for a pedagogical introduction.

The operator construction of the $1/N_c$ expansion was later extended to study
the masses of the negative parity $L=1$ excited baryons \cite{Goity:1996hk}
\cite{Pirjol:1997bp} \cite{Pirjol:1997sr} \cite{Carlson:1998vx}
\cite{Schat:2001xr} \cite{Goity:2002pu} with great success. The strong and
electromagnetic decays of these states belonging to the $[\mathbf{70},1^-]$ multiplet,
as well as the masses and decays of baryon resonances in other spin-flavor
multiplets were also studied in the $1/N_c$ expansion (see
\cite{Matagne:2014lla} for a recent review, and references therein), establishing a
comprehensive framework to study the phenomenology of excited baryons at the
physical value $N_c=3$.  Lattice studies of excited baryons offer further 
tests of the $1/N_c$ expansion by providing 
predictions for all states in a spin-flavor multiplet and probing the dependence of 
observables  on the 
quark masses \cite{Jenkins:2009wv} \cite{Fernando:2014dna}. They are even starting to explore 
the mass spectrum for $N_c$ values larger than three  \cite{Cordon:2014sda}
\cite{DeGrand:2016pur}.

It is important to note that the  classification scheme for
baryon resonances based on irreducible representations (irreps) 
of $SU(6)\times O(3)$, grouped into excitation bands $N=0,1,2,3,...$, is based on the
quark model and cannot be justified from QCD. 
Physical states appear as admixtures of different $SU(6)\times O(3)$
irreps, something known as {\it configuration mixing},
which so far has been neglected in phenomenological studies that use the
$1/N_c$ expansion. 

In the large $N_c$ limit a different symmetry structure is present at leading
order and states are classified according to irreducible representations 
of the contracted symmetry $SU(2F)_c$, dubbed as {\em towers} and
labeled by $K$, leading to degeneracies in the mass spectrum of excited baryons
\cite{Pirjol:1997bp,Pirjol:1997sr}.  These predicted degeneracies in the large
$N_c$ limit have been verified by explicit calculations, first in
\cite{Pirjol:2003ye,Cohen:2003tb} for the nonstrange $[{\bf 70},1^-]$ states that
constitute the $N=1$ band and later also for the nonstrange $[\mathbf{70},3^-]$ states of the $N=3$
band \cite{Matagne:2011sn}, but up to now the effect of configuration mixing,
which is not $N_c$ suppressed \cite{Goity:2004pw,Goity:2005gs}, 
has been neglected in an explicit calculation of the spectrum 
using the effective quark operators. 

Here we consider the entire space of states spanned by the $N=2$ states $[{\bf
70},L^+]$ and $[{\bf 56},L^+]$, with $L=0,2$.  We restrict ourselves to 
the nonstrange states and show by an explicit
calculation that configuration mixing preserves the tower structure of the mass
spectrum in the large $N_c$ limit.  We also study the mixing pattern, which in the
basis of $K$ states turns out to be much simpler than what would have been
expected starting from the spin-flavor basis of the quark model. 

In order to perform the calculation including the leading order effects of
configuration mixing we need to extend the construction of the leading order
mass operator. To allow for $L \rightarrow L'$ transitions, we will introduce a
generic spatial operator $\xi$ in place of the angular momentum operator $\ell$ used
in previous studies  \cite{Carlson:1998vx}. This is similar to the construction of the transition
operators for the decay processes \cite{Goity:2004ss}, as we need the most
general operators that mix the $SU(6)\times O(3)$ multiplets to leading order. 

In this work we present the most general form of the leading order mass operator that
incorporates configuration mixing and the explicit expressions obtained for its
matrix elements between states of the $N=2$ band. Our results provide
important consistency checks on the contracted symmetry predictions for the
masses and mixings in the large $N_c$ limit, and the correctness of the usual
construction of the effective mass operators in terms of core and excited quark
operators \cite{Carlson:1998vx}.

The paper is organized as follows: In Sec.~\ref{sec:thestates} we present the
states, in Sec.~\ref{sec:massoperator} we present the effective mass operator, in
Sec.~\ref{sec:massmat}  we give the explicit form of the mass matrices for
$I=1/2$ and $I=3/2$ states. In Sec.~\ref{sec:Discussion} we discuss the spectrum
and the mixing pattern we obtain. We finally conclude in
Sec.~\ref{sec:conclusions}. In App.~\ref{app:A} we give the general expressions
for calculating the matrix elements of the leading mass operators and in
App.~\ref{app:B} we give the explicit expressions of the matrix elements for
arbitrary $N_c$.

\section{The states}
\label{sec:thestates}

The positive parity, orbitally excited modes of an $N_c=3$ three quark system
differ from each other in their orbital angular momentum $L$ and their
behaviour under permutations. The spin-flavor symmetry $SU(6)$ is a useful
classification scheme for these states. Each type of permutational symmetry
of three objects, symmetric $(S)$, mixed symmetric $(MS)$ and antisymmetric 
$(A)$ corresponds to an $SU(6)$ multiplet,   
$\mathbf{56}$, $\mathbf{70}$ and $\mathbf{20}$, respectively. For the
nonstrange members of these multiplets, the spin-flavor symmetry is only
broken by the spin-dependent interactions. Here and in what follows 
we will concentrate on the nonstrange members of each $SU(6)\times O(3)$ multiplet.
In the case of harmonic forces between
quarks the five $SU(6)\times O(3)$
multiplets  $\mathbf{56'}$ ($L=0,2$), $\mathbf{70}$ ($L=0,2$) and $\mathbf{20}$ ($L=1$) 
are degenerate and constitute the harmonic oscillator $N=2$ band.  
We do not consider the antisymmetric states in this work, as their relevance for 
the observed physical states at $N_c=3$ is not clear yet.
In the generalization of the lowest-energy multiplets to arbitrary $N_c$ the
additional $N_c - 3$ quarks are taken in a
completely symmetric spin-flavor combination.

In order to identify the nonstrange $I=1/2$ and $I=3/2$ physical states ($N$ and $\Delta$) with 
the large $N_c$ states, it is useful to keep in mind the usual 
$SU(6)_{spin-flavor} \supset SU(3)_{flavor}\times SU(2)_{spin}$ decomposition that labels the  
$N_c=3$ states
\begin{equation}\label{states3}
\begin{array}{lll}
S^{N_c=3} & : \mathbf{56} = {^4}\mathbf{10} + {^2}\mathbf{8}   & \supset {^4}\Delta + {^2}N \ , \\
MS^{N_c=3} &  : \mathbf{70} = {^2}\mathbf{1}  + {^2}\mathbf{10} + {^2}\mathbf{8} + {^4}\mathbf{8}   & \supset {^2}\Delta + {^2}N + {^4}N \ , 
\end{array} 
\end{equation}
where  ${^{2S+1}}N,{^{2S+1}}\Delta$, with $S$ the quark spin of the state.
It should always be clear if $S$ stands for the spin, or if it denotes the symmetric irrep of the 
permutation group. 
To count the nonstrange  $I=1/2$ and $I=3/2$ states contained in the $S$, $MS$ irreducible representations 
of the permutation group for arbitrary 
$N_c$ it is helpful to recall that for the symmetric representation the spin and 
isospin of the states are related by $S = I $, while for the 
mixed symmetric irrep the spin is obtained from the vector sum $\mathbf{S} = \mathbf{I}+\mathbf{1}$, so that 
\begin{equation}\label{statesN}
\begin{array}{ll}
S &   \supset {^4}\Delta +  {^2}N  \, , \\
MS &   \supset {^2}\Delta + \underbrace{{^4}\Delta + {^6}\Delta}_{N_c \ge 5} +  {^2}N + {^4}N \ .
\end{array} 
\end{equation}
We see that in $MS$ there appear $\Delta$ states with higher spin that are absent in $MS^{N_c=3}$.
These are ghost states that decouple from the physical states in the $N_c \rightarrow 3$ limit, as 
has been noticed in  
\cite{Pirjol:2003ye}
\cite{Cohen:2006en} 
and we will also see explicitly here when discussing the matrix elements of App.~\ref{app:B}. 
After coupling with the orbital angular momentum $L=0,2$  we obtain the 
$S_L, MS_L$ states 
\begin{equation}\label{states}
\begin{array}{lll}
S'_{0} & \supset \, & N_{1/2},\Delta'_{3/2} \ , \\                                                               
MS_{0} & \supset \, &N_{1/2},N'_{3/2},\Delta_{1/2},\underbrace{\Delta'_{3/2},\Delta''_{5/2}}_{N_c \ge 5}
 \ , \\                           
S_{2} & \supset \, &N_{3/2}, N_{5/2},\Delta'_{1/2}, \Delta'_{3/2}, \Delta'_{5/2}, \Delta'_{7/2}  \ , \\          
MS_{2} & \supset \, &N_{3/2}, N_{5/2}, N'_{1/2}, N'_{3/2}, N'_{5/2}, N'_{7/2},\Delta_{3/2}, \Delta_{5/2},  \\ 
&&\underbrace{\Delta'_{1/2}, \Delta'_{3/2}, \Delta'_{5/2}, \Delta'_{7/2}, 
\Delta''_{1/2}, \Delta''_{3/2}, \Delta''_{5/2}, \Delta''_{7/2}, \Delta''_{9/2}}_{N_c \ge 5}
 \ , 
\end{array} 
\end{equation}
where we indicate $J$, the total spin of the nonstrange states $N_J,\Delta_J$,  
as given by the vector sum $\mathbf{J}=\mathbf{S}+\mathbf{L}$. The 
primes indicate  the different values of quark spin $S$, e.g. $N_J,N'_J$ correspond to ${^{2S+1}}N_J$
with $S=1/2,3/2$, respectively, while  $\Delta_J,\Delta'_J,\Delta''_J$  correspond to ${^{2S+1}}\Delta_J$
with $S=1/2,3/2,5/2$, respectively. 

In Eq.~(\ref{states}) we show all the states that we will consider to compute the mass spectrum. There are 
30 $I=1/2,3/2$ states distributed as 11 $N$-states and 19 $\Delta$-states, where 11 of the $\Delta$-states are
ghost states. That reduces the number of physical states at $N_c=3$ to 19 states.

The representations $S'_{0}, MS_{0}, S_{2}, MS_{2}$ reduce in the $N_c=3$ limit  
to  the $[\mathbf{56'},0^+]$,$[\mathbf{56},2^+]$, $[\mathbf{70},0^+] $
and $[\mathbf {70},2^+]$ quark model $SU(6) \times O(3)$ multiplets. 
We use the prime on $S'_{0}$ to distinguish it from the ground state baryons $S_{0}$, usually labeled as 
$[\mathbf {56},0^+]$ in the $N_c=3$ limit.

In a quark model calculation all states with same $I,J$ mix, giving rise to large mixing matrices 
and a complex pattern, see e.g. Ref.~\cite{Isgur:1978wd}. 
In the large $N_c$ limit the mixing pattern is much simpler, something that only becomes 
clear after classifying the states in a different way, according to the irreps 
of the contracted spin-flavor symmetry $SU(4)_c$. These irreps are labeled by $K$, 
which relates $J$ and $I$ as $\mathbf{J}=\mathbf{I}+\mathbf{K}$, so that $K=L$ in the symmetric irreps and $\mathbf{K}=\mathbf{L+1}$ 
in the mixed symmetric ones \cite{Pirjol:1997bp}\cite{Pirjol:1997sr}. 
In contrast to the spin-flavor states for arbitrary $N_c$ of Eq.~(\ref{states}), 
the 30 large $N_c$ states 
carry an additional label $K$ and can be grouped as ``tower states" 
\begin{equation}\label{Kstates}
\begin{array}{lcl}
K=0 &:\, &N_{1/2}, \Delta_{3/2}, \cdots \\
K=1 &:\, &N_{1/2}, N_{3/2},\Delta_{1/2},\Delta_{3/2},\Delta_{5/2}, \cdots       \\
K=1'&:\, &N_{1/2}, N_{3/2},\Delta_{1/2},\Delta_{3/2},\Delta_{5/2}, \cdots       \\
K=2 &:\, &N_{3/2}, N_{5/2},\Delta_{1/2}, \Delta_{3/2}, \Delta_{5/2}, \Delta_{7/2}, \cdots \\
K=2'&:\, &N_{3/2}, N_{5/2},\Delta_{1/2}, \Delta_{3/2}, \Delta_{5/2}, \Delta_{7/2}, \cdots \\
K=3 &:\, &N_{5/2}, N_{7/2},\Delta_{3/2}, \Delta_{5/2}, \Delta_{7/2}, \Delta_{9/2}, \cdots 
\end{array}
\end{equation}
where each tower state is in general in an admixture of the spin-flavor states shown in Eq.~(\ref{states}).  
The nonstrange states in $S'_0$ belong to a $K=0$ tower, the ones in $MS_0$ appear in the decomposition
of $K=1$ states, the ones in $S_2$ contribute to  
$K=2$ states and finally the $MS_2$ states appear in the decomposition of $K=1,2,3$ states. 
As we will show by an explicit calculation in Sec.~\ref{sec:Discussion}, in the $K$ basis only states with the same 
$K$ label mix.
 This implies that in the large $N_c$ limit 
the tower structure that was first found by performing explicit 
calculations within a single spin-flavor irrep \cite{Pirjol:2003ye}\cite{Cohen:2003tb}\cite{Matagne:2011sn}
is also preserved when including 
the effect of configuration mixing.

In the next Section we will present the simple leading order in $N_c$ mass operator 
from where this mixing pattern follows.

\section{The mass operator}
\label{sec:massoperator}

The leading order mass operator needed for our calculation is obtained by slightly generalizing the 
construction of Ref.~\cite{Carlson:1998vx} as follows. As explained in detail in  
Ref.~\cite{Carlson:1998vx}, 
the large $N_c$ states can be constructed as product states of a
symmetric core of $N_c-1$ quarks and an  $N_c$-th quark in a proper linear 
combination, so that the  $S^{N_c}$ and $MS^{N_c}$ irreps
with the desired permutation properties are obtained. 
The operators contributing to the mass operator at different 
orders in $1/N_c $ can then be constructed
from the $SU(6)$ generators acting on the symmetric core and on the quark
that was singled out, and  the orbital angular momentum operator $\ell$. Here we will replace 
the orbital angular momentum operator by a generic spatial operator $\xi$
 to allow for configuration mixing of spin-flavor 
representations with different $L$.
This is similar to the construction of the effective operators for decay processes
\cite{Carone:1994tu}
\cite{Goity:2004ss}, 
where we also needed to describe transitions between states 
with different $L$.  

The leading order Hamiltonian including operators up to order ${\cal O}(N_c^0)$
that will correctly describe all possible  mixings in our configuration space,  
spanned by the states given in Eq.~(\ref{states}), has then the following form 
\begin{equation}\label{massop}
 H = c_1^{\mathbf{R}} \openone + c_2^{\mathbf{R},\mathbf{R'}} \xi \cdot s 
+ c_3^{\mathbf{R},\mathbf{R'}} \frac{1}{N_c} \xi^{(2)} \cdot g \cdot G_c
 + {\cal O}(1/N_c) \ ,  
\end{equation}
where 
$\xi^{(2)ij}=\frac{1}{2}\{\xi^i,\xi^j\} -
\frac{\xi^2}{3}\delta^{ij}$ 
and
$\mathbf{R}$, $\mathbf{R'}$ stand for the $S_N \times O(3)$ irreps  $S_L,MS_L$. 
The coefficients $c_{1,2,3}^{\mathbf{R},\mathbf{R'}}$ are order $ {\cal O}(N_c^0) $ and 
encode the details of the spatial wave function, as 
has been shown explicitly within a single spin-flavor representation 
by different matching calculations  
\cite{Collins:1998ny} \cite{Pirjol:2007ed} \cite{Galeta:2009pn} 
\cite{Pirjol:2010th} \cite{Willemyns:2015hgy}.
 They 
take different values on different spin-flavor irreps $\mathbf{R}$. In our 
case we also have off-diagonal matrix elements, so that the coefficients depend
on both the $\mathbf{R}$ and $\mathbf{R'}$ irreps that are mixing. For 
the diagonal matrix elements we have 
$\mathbf{R}=\mathbf{R'}$ and we use the notation $c_i^{\mathbf{R}} = c_i^{\mathbf{R},\mathbf{R}} $.
The unit operator only contributes to diagonal matrix elements. 

The general expressions for the matrix elements of the operators in Eq.~(\ref{massop}) are  
given in App.~\ref{app:A}  following closely the notation of Ref.~\cite{Carlson:1998vx}. 

\section{The mass matrices}
\label{sec:massmat}
In this Section we present the explicit form of the mass matrices we obtain by computing the matrix elements of
 Eq.~(\ref{massop}),  after expanding in $1/N_c$ and taking the large $N_c$ limit. 
The matrix elements for arbitrary $N_c$ are given in App.~\ref{app:B}.
We also give the expressions for the corresponding eigenvalues and eigenstates. 
The reader can skip through this Section and continue reading Sec.~\ref{sec:Discussion}, 
where we work out an explicit example that better illustrates the relevant points of 
the discussion.

It should be noted that in the coefficients 
$c_1^{\mathbf{R}},  c_2^{\mathbf{R},\mathbf{R'}},  c_3^{\mathbf{R},\mathbf{R'}}$ we 
absorb common group theoretical numerical factors and the reduced matrix elements of the $\xi$ operator,
as appearing in Eqs.~(\ref{o2xi}, \ref{o3xi}) or more explicitly, as in Tables~\ref{table12}, 
\ref{table3212}, \ref{table3232} and \ref{table3252} of 
App.~\ref{app:B}. 
We do not distinguish them from the original 
$c_1^{\mathbf{R}},  c_2^{\mathbf{R},\mathbf{R'}},  c_3^{\mathbf{R},\mathbf{R'}}$ 
appearing in Eq.~(\ref{massop}) to keep the notation simple.   

\subsection{The $I=1/2$ states}
The $I=1/2$ states with the same $J$ can mix among each other. We have three $J=1/2$, four 
$J=3/2$, three $J=5/2$ and one $J=7/2$ states. Their mass matrices are given below.
The 11 mass eigenvalues for the large $N_c$, $I=1/2$ states will be labeled as $ m_{N_{J}^{K}} $ 
and their degeneracies will be discussed in the next Section.
Table \ref{table12} in App.~\ref{app:B} shows all the matrix elements for the nucleons at finite $N_c$. 

In the large $N_c$ limit the mass matrix  for the $N_{1/2}$ states 
in the \{${^2}N_{1/2}^{S_0'}$, ${^2}N_{1/2}^{MS_0}$, ${^4}N_{1/2}^{MS_2}$\} basis is  
\begin{eqnarray}
M_{N_{1/2}}
&=&
  \left(
\begin{array}{ccc}
 c_1^{S_0'} N_c& 0 		& 0 \\
	      & c_1^{MS_0} N_c	& \sqrt{2} c_3^{MS_0 MS_2} \\
	      & 		& c_1^{MS_2} N_c-\frac{3}{2} c_2^{MS_2}-c_3^{MS_2}\\
\end{array}
\right) \ .
 \end{eqnarray}
The $M_{N_{1/2}}$ mass matrix and all mass matrices that follow are symmetric. We only show the upper right half 
of the matrix to keep the expressions more readable. 
The eigenvalues are $m_{N_{1/2}^{K=0 \;}},  m_{N_{1/2}^{K=1' \;}} $ and $m_{N_{1/2}^{K=1}} $.
Their explicit dependence on the $c_i^{\mathbf{R},\mathbf{R'}}$ coefficients is given 
in the next Section. The corresponding eigenstates are
\begin{eqnarray}
\left(
\begin{array}{c}
N_{1/2}^{K=0} \\
N_{1/2}^{K=1'} \\
N_{1/2}^{K=1} 
\end{array}
\right)
&=&
\left(
\begin{array}{ccc}
1 & 0		&  0 \\
0 & 1		& -\eta_{MS_0} \\
0 & \eta_{MS_0}	& 1 
\end{array}
\right) \ , 
\end{eqnarray}
where each row vector on the right hand side indicates the composition of the eigenstate, e.g. 
for the second eigenstate we have
\begin{eqnarray}
| N_{1/2}^{K=1'} \rangle &=& | {^2}N_{1/2}^{MS_0}\rangle   -\eta_{MS_0} | {^4}N_{1/2}^{MS_2}\rangle  \ , 
\end{eqnarray}
where $\eta_{MS_0}$ can be expressed in terms of the $c_i^{\mathbf{R},\mathbf{R'}}$ coefficients.
We will use this  matrix notation to show the composition of the eigenstates throughout the rest of the
paper. This is very convenient to make the mixing pattern manifest.
In the limit of no mixing ($\eta_\mathbf{R}=0$) the eigenstates are normalized. 
In the general case there is still a normalization factor of ${\cal N}=\frac{1}{\sqrt{1+\eta_\mathbf{R}^2}}$ 
that has to be taken into account.

The mass matrix for $N_{3/2}$ states in the large $N_c$ limit is given in 
the  \{${^4}N_{3/2}^{MS_0}$, ${^2}N_{3/2}^{S_2}$, ${^2}N_{3/2}^{MS_2}$, ${^4}N_{3/2}^{MS_2}$\} basis 
by 
\begin{eqnarray}
M_{N_{3/2}}
&=&
  \left(
\begin{array}{cccc}
 c_1^{MS_0} N_c	& 0 		& -c_3^{MS_0 MS_2} 		& -c_3^{MS_0 MS_2} \\
		& c_1^{S_2} N_c	& c_2^{S_2 MS_2} 		& -c_2^{S_2 MS_2} \\
		& 		& c_1^{MS_2} N_c-c_2^{MS_2} 	& -\frac{1}{2} c_2^{MS_2}- c_3^{MS_2} \\
		& 		& 				& c_1^{MS_2} N_c-c_2^{MS_2} \\
\end{array}
\right) \ .
 \end{eqnarray}
We denote the eigenvalues as $m_{N_{3/2}^{K=1' \;}},   m_{N_{3/2}^{K=2' \;}},  m_{N_{3/2}^{K=1 \;}}, $ and $m_{N_{3/2}^{K=2}} $.
The eigenstates of this matrix are
\begin{eqnarray}
\left(
\begin{array}{c}
N_{3/2}^{K=1'} \\
N_{3/2}^{K=2'} \\
N_{3/2}^{K=1} \\
N_{3/2}^{K=2}
\end{array}
\right)
&=&
\left(
\begin{array}{cccc}
1 		& 0 		& \frac{\sqrt{2}}{2}\eta_{MS_0}	& \frac{\sqrt{2}}{2}\eta_{MS_0} \\
0 		& 1		& -\frac{\sqrt{2}}{2}\eta_{S_2}	& \frac{\sqrt{2}}{2}\eta_{S_2} \\
-\eta_{MS_0} 	& 0		& \frac{\sqrt{2}}{2}		& \frac{\sqrt{2}}{2} \\
0 		& -\eta_{S_2} 	& -\frac{\sqrt{2}}{2}		& \frac{\sqrt{2}}{2} 
\end{array}
\right) \ .
\end{eqnarray}

For the $N_{5/2}$ states in the 
\{${^2}N_{5/2}^{S_2}$, ${^2}N_{5/2}^{MS_2}$, ${^4}N_{5/2}^{MS_2}$\} basis 
we obtain
\begin{eqnarray}
M_{N_{5/2}}
&=&
\left(
\begin{array}{ccc}
 c_1^{S_2} N_c 	& -\frac{2}{3} c_2^{S_2 MS_2} 		& -\frac{\sqrt{14}}{3} c_2^{S_2 MS_2} \\
		& c_1^{MS_2} N_c+\frac{2}{3}c_2^{MS_2} 	& -\frac{\sqrt{14}}{6}c_2^{MS_2} + \frac{\sqrt{14}}{6} c_3^{MS_2} \\
		& 					& c_1^{MS_2} N_c-\frac{1}{6}c_2^{MS_2}+\frac{5}{7}c_3^{MS_2}\\
\end{array}
\right) \ , 
 \end{eqnarray}
with eigenvalues 
$m_{N_{5/2}^{K=2' \;}},   m_{N_{5/2}^{K=2\;}}$ and $m_{N_{5/2}^{K=3 \;}}, $
and eigenstates 
\begin{eqnarray}
\left(
\begin{array}{c}
N_{5/2}^{K=2'} \\
N_{5/2}^{K=2} \\
N_{5/2}^{K=3}
\end{array}
\right)
&=&
\left(
\begin{array}{ccc}
1		& \frac{\sqrt{2}}{3}\eta_{S_2}	& \frac{\sqrt{7}}{3}\eta_{S_2} \\
-\eta_{S_2}	& \frac{\sqrt{2}}{3}		& \frac{\sqrt{7}}{3} \\
0		& -\frac{\sqrt{7}}{3} 		& \frac{\sqrt{2}}{3}
\end{array}
\right) \ .
\end{eqnarray}

Finally, the matrix element for the $N_{7/2}$ large $N_c$ state in $MS_2$ is
 \begin{eqnarray}
    m_{N_{7/2}^{K=3}}&=&c_1^{MS_2} N_c+c_2^{MS_2}-\frac{2}{7} c_3^{MS_2} \ .
 \end{eqnarray}

\subsection{The $I=3/2$ states}
The $I=3/2$ states with the same $J$ can mix among each other. We have four $J=1/2$, six 
$J=3/2$, five $J=5/2$, three $J=7/2$ and one $J=9/2$ states. Their mass matrices are given below.
The 19 mass eigenvalues for the large $N_c$, $I=3/2$ states will be labeled as $ m_{\Delta_{J}^{K}} $ 
and their degeneracies will be discussed in the next Section.
  
Table \ref{table3212} in App.~\ref{app:B}  shows the matrix elements for $\Delta_{1/2}$ at finite $N_c$.
In the large $N_c$ limit the mass matrix for the $\Delta_{1/2}$ states in the 
\{${^2}\Delta_{1/2}^{MS_0}$, ${^4}\Delta_{1/2}^{S_2}$, ${^4}\Delta_{1/2}^{MS_2}$, ${^6}\Delta_{1/2}^{MS_2}$\}
basis is 
\begin{eqnarray}
 M_{\Delta_{1/2}}
 &=&
  \left(
\begin{array}{cccc}
 c_1^{MS_0} N_c	& 0 		& \frac{1}{\sqrt{5}}c_3^{MS_0 MS_2} 				& \frac{3}{\sqrt{5}}c_3^{MS_0 MS_2} \\
 		& c_1^{S_2} N_c	& \frac{3}{\sqrt{5}}c_2^{S_2 MS_2} 				& -\frac{1}{\sqrt{5}}c_2^{S_2 MS_2} \\
		&		& c_1^{MS_2} N_c-\frac{3}{5}c_2^{MS_2}+\frac{4}{5}c_3^{MS_2}	& -\frac{3}{10}c_2^{MS_2}-\frac{3}{5}c_3^{MS_2} \\
		&		&								& c_1^{MS_2}N_c-\frac{7}{5} c_2^{MS_2}-\frac{4}{5}c_3^{MS_2}
\end{array}
\right) \ , 
 \end{eqnarray}
with eigenvalues $m_{\Delta_{1/2}^{K=1'}},m_{\Delta_{1/2}^{K=2'}},m_{\Delta_{1/2}^{K=1}}$ and $m_{\Delta_{1/2}^{K=2}}$.
The corresponding eigenstates are
\begin{eqnarray}
\left(
\begin{array}{c}
\Delta_{1/2}^{K=1'} \\
\Delta_{1/2}^{K=2'} \\
\Delta_{1/2}^{K=1} \\
\Delta_{1/2}^{K=2}
\end{array}
\right)
&=&
\left(
\begin{array}{cccc}
1				& 0						& -\frac{1}{2}\sqrt{\frac{2}{5}}\eta_{MS_0} 	& -\frac{3}{2}\sqrt{\frac{2}{5}}\eta_{MS_0} \\
0				& 1						& -\frac{3}{2}\sqrt{\frac{2}{5}}\eta_{S_2}	& \frac{1}{2}\sqrt{\frac{2}{5}}\eta_{S_2} \\
-\eta_{MS_0}			& 0						& -\frac{1}{2}\sqrt{\frac{2}{5}}		& -\frac{3}{2}\sqrt{\frac{2}{5}} \\
0				& -\eta_{S_2}					& -\frac{3}{2}\sqrt{\frac{2}{5}}		& \frac{1}{2}\sqrt{\frac{2}{5}}
\end{array}
\right).
\end{eqnarray}
Table \ref{table3232} shows the matrix elements for $\Delta_{3/2}$ at finite $N_c$.
For the $\Delta_{3/2}$ states in the 
\{
${^4}\Delta_{3/2}^{S'_0}$, 
${^4}\Delta_{3/2}^{MS_0}$, 
${^4}\Delta_{3/2}^{S_2}$, 
${^2}\Delta_{3/2}^{MS_2}$, 
${^4}\Delta_{3/2}^{MS_2}$, 
${^6}\Delta_{3/2}^{MS_2}$\}
basis we obtain
\begin{eqnarray}
M_{\Delta_{3/2}}
&=&\left(
\begin{array}{cccccc}
 c_1^{S'_0} N_c	& 0 			& 0 					& 0 					& 0 								& 0 \\
 		& c_1^{MS_0} N_c	& 0 					& -\frac{1}{\sqrt{10}}c_3^{MS_0 MS_2} 	& \frac{4}{5}c_3^{MS_0 MS_2} 					& \frac{3}{5} \sqrt{\frac{7}{2}} c_3^{MS_0 MS_2} \\
  		&  			& c_1^{S_2} N_c				& -\frac{1}{\sqrt{2}}c_2^{S_2 MS_2} 	& \frac{2}{\sqrt{5}}c_2^{S_2 MS_2} 				& -\sqrt{\frac{7}{10}} c_2^{S_2 MS_2} \\
  		& 			& 					& c_1^{MS_2} N_c+\frac{1}{2}c_2^{MS_2}	& \frac{5}{2\sqrt{10}} c_2^{MS_2}-\frac{1}{\sqrt{10}} c_3^{MS_2}& \frac{3}{\sqrt{35}}c_3^{MS_2} \\
  		& 			& 					& 					& c_1^{MS_2} N_c-\frac{2}{5}c_2^{MS_2}				& -\frac{3}{10}\sqrt{\frac{7}{2}} c_2^{MS_2}-\frac{3}{7}\sqrt{\frac{7}{2}} c_3^{MS_2} \\
  		&			& 					& 					& 								& c_1^{MS_2} N_c -\frac{11}{10} c_2^{MS_2}-\frac{2}{7} c_3^{MS_2} 	
\end{array}
\right) \ ,  \nonumber \\
\label{example32}
\end{eqnarray}
with eigenvalues 
$m_{\Delta_{3/2}^{K=0}}, m_{\Delta_{3/2}^{K=1'}}, m_{\Delta_{3/2}^{K=2'}}, m_{\Delta_{3/2}^{K=1}}$, 
$m_{\Delta_{3/2}^{K=2}}$ and $m_{\Delta_{3/2}^{K=3}}$
and eigenstates
\begin{eqnarray}
\label{mD4}
\left(
\begin{array}{c}
\Delta_{3/2}^{K=0} \\
\Delta_{3/2}^{K=1'} \\
\Delta_{3/2}^{K=2'}  \\
\Delta_{3/2}^{K=1} \\
\Delta_{3/2}^{K=2}  \\
\Delta_{3/2}^{K=3} \\
\end{array}
\right)
&=&
\left(
\begin{array}{cccccc}
1 & 0		& 0		& 0				& 0					& 0 \\
0 & 1 		& 0		& \frac{1}{2\sqrt{5}}\eta_{MS_0}	& -\frac{2\sqrt{2}}{5}\eta_{MS_0}	& -\frac{3\sqrt{7}}{10}\eta_{MS_0} \\
0 & 0		& 1		& \frac{1}{2}\eta_{S_2}		& -\sqrt{\frac{2}{5}}\eta_{S_2}		& \frac{1}{2}\sqrt{\frac{7}{5}}\eta_{S_2} \\
0 & -\eta_{MS_0}	& 0 		& \frac{1}{2\sqrt{5}} 		& -\frac{2\sqrt{2}}{5}			& -\frac{3\sqrt{7}}{10} \\
0 & 0		& -\eta_{S_2}	& \frac{1}{2}			& -\sqrt{\frac{2}{5}}			& \frac{1}{2}\sqrt{\frac{7}{5}} \\
0 & 0		& 0		& -\sqrt{\frac{7}{10}}		& -\frac{\sqrt{7}}{5}			& \frac{1}{5\sqrt{2}}
\end{array}
\right).
\end{eqnarray}
 
Table \ref{table3252} in App.~\ref{app:B} shows the matrix elements for $\Delta_{5/2}$, $\Delta_{7/2}$  and $\Delta_{9/2}$ at finite $N_c$.
For the $\Delta_{5/2}$ states in the 
\{
${^6}\Delta_{5/2}^{MS_0}$, 
${^4}\Delta_{5/2}^{S_2}$, 
${^2}\Delta_{5/2}^{MS_2}$, 
${^4}\Delta_{5/2}^{MS_2}$, 
${^6}\Delta_{5/2}^{MS_2}$\}
basis we have  
 \begin{eqnarray}
M_{\Delta_{5/2}}
 &=&
\left(
\begin{array}{ccccc}
c_1^{MS_0} N_c 	& 0 		& \sqrt{\frac{3}{5}} c_3^{MS_0MS_2} 	& -\frac{1}{5} \sqrt{21} c_3^{MS_0MS_2}							& -\frac{1}{5} \sqrt{14} c_3^{MS_0MS_2} \\
		& c_1^{S_2} N_c	& -\frac{1}{3} \sqrt{7} c_2^{S_2MS_2}	& \frac{1}{3 \sqrt{5}}c_2^{S_2MS_2} 							& -\sqrt{\frac{6}{5}} c_2^{S_2MS_2} \\
		& 		& c_1^{MS_2} N_c-\frac{1}{3}c_2^{MS_2}	& \frac{5}{6} \sqrt{\frac{7}{5}} c_2^{MS_2}+\frac{1}{7}\sqrt{\frac{7}{5}} c_3^{MS_2}	& 2 \sqrt{\frac{6}{35}} c_3^{MS_2} \\
		& 		& 					& c_1^{MS_2} N_c-\frac{1}{15}c_2^{MS_2}-\frac{4}{7} c_3^{MS_2}				& -\frac{3}{5}\sqrt{\frac{3}{2}}c_2^{MS_2}-\frac{2}{7}\sqrt{\frac{3}{2}}c_3^{MS_2} \\
		& 		& 					& 											& c_1^{MS_2} N_c -\frac{3}{5} c_2^{MS_2}+\frac{2}{7} c_3^{MS_2}
\end{array}
\right) \ ,  \nonumber \\
 \end{eqnarray}
with eigenvalues 
$m_{\Delta_{5/2}^{K=1'}}, m_{\Delta_{5/2}^{K=2'}}, m_{\Delta_{5/2}^{K=1}}, m_{\Delta_{5/2}^{K=2}}$, 
and $m_{\Delta_{5/2}^{K=3}}$
and eigenstates  
\begin{eqnarray}
\left(
\begin{array}{c}
\Delta_{5/2}^{K=1'} \\
\Delta_{5/2}^{K=2'} \\
\Delta_{5/2}^{K=1} \\
\Delta_{5/2}^{K=2} \\
\Delta_{5/2}^{K=3} 
\end{array}
\right)
&=&
\left(
\begin{array}{ccccc}
1		& 0					& -\frac{1}{2}\sqrt{\frac{6}{5}}\eta_{MS_0}	& \frac{1}{5}\sqrt{\frac{21}{2}}\eta_{MS_0} 	& \frac{\sqrt{7}}{5}\eta_{MS_0} \\
0		& 1					& \frac{1}{3}\sqrt{\frac{7}{2}}\eta_{S_2}	& -\frac{1}{6}\sqrt{\frac{2}{5}}\eta_{S_2} 	& \sqrt{\frac{3}{5}}\eta_{S_2} \\
-\eta_{MS_0}	& 0					& -\frac{1}{2}\sqrt{\frac{6}{5}}		& \frac{1}{5}\sqrt{\frac{21}{2}}		& \frac{\sqrt{7}}{5} \\
0		& -\eta_{S_2}				& \frac{1}{3}\sqrt{\frac{7}{2}}			& -\frac{1}{6}\sqrt{\frac{2}{5}}		& \sqrt{\frac{3}{5}} \\
0 		& 0					& -\frac{1}{3}\sqrt{\frac{14}{5}}		& -\frac{8\sqrt{2}}{15}				& \frac{\sqrt{3}}{5} \\
\end{array}
\right) \ .
\end{eqnarray}
The matrix for the $\Delta_{7/2}$ states in the large $N_c$ limit in the 
\{
${^4}\Delta_{7/2}^{S_2}$, 
${^4}\Delta_{7/2}^{MS_2}$, 
${^6}\Delta_{7/2}^{MS_2}$\}
basis is
\begin{eqnarray}
M_{\Delta_{7/2}}=
\left(
\begin{array}{ccc}
 c_1^{S_2} N_c 	& -\frac{2}{\sqrt{5}}c_2^{S_2MS_2} 					& -\sqrt{\frac{6}{5}} c_2^{S_2MS_2} 								\\
		& c_1^{MS_2} N_c+\frac{2}{5} c_2^{MS_2}+\frac{8}{35}c_3^{MS_2} 		& -\frac{3}{5} \sqrt{\frac{3}{2}}c_2^{MS_2}+\frac{18}{35}\sqrt{\frac{3}{2}} c_3^{MS_2} 	\\
		& 											& c_1^{MS_2} N_c +\frac{1}{10} c_2^{MS_2}+\frac{17}{35}c_3^{MS_2} 			\\
\end{array}
\right) \ ,  
 \end{eqnarray}
with eigenvalues 
$m_{\Delta_{7/2}^{K=2'}}, m_{\Delta_{7/2}^{K=2}}$, 
and $m_{\Delta_{7/2}^{K=3}}$, and 
with eigenstates
\begin{eqnarray}
\left(
\begin{array}{c}
\Delta_{7/2}^{K=2'} \\
\Delta_{7/2}^{K=2} \\
\Delta_{7/2}^{K=3} 
\end{array}
\right)
&=&
\left(
\begin{array}{ccc}
1		& \sqrt{\frac{2}{5}}\eta_{S_2}	& \sqrt{\frac{3}{5}}\eta_{S_2}\\
-\eta_{S_2}	& \sqrt{\frac{2}{5}}		& \sqrt{\frac{3}{5}} \\
0		& -\sqrt{\frac{3}{5}}		& \sqrt{\frac{2}{5}}
\end{array}
\right).
\end{eqnarray}

Finally, the matrix element for the large $N_c$ state $\Delta_{9/2}$ in $MS_2$  is
\begin{eqnarray}
   m_{\Delta_{9/2}^{K=3}}&=&c_1^{MS_2} N_c + c_2^{MS_2}-\frac{2}{7}  c_3^{MS_2} \ .
\end{eqnarray}

\section{The large $N_c$ spectrum}
\label{sec:Discussion}

The diagonalization of the $I=1/2$ and $I=3/2$ mass matrices presented in the 
previous Section leads to 30 mass eigenvalues. In the large $N_c$ limit we 
find as a result of our explicit calculation the remarkable result that 
the masses assume only six different values, leading to a highly 
degenerate spectrum. The 11 $I=1/2$ and 19 $I=3/2$ masses are grouped in 
six energy levels $m_K$ as follows  
\begin{eqnarray}
m_{0 \; } &=& m_{N_{1/2}^{K=0 \;}}=m_{\Delta_{3/2}^{K=0}} \ , \nonumber \\  
m_{1'} &=& m_{N_{1/2}^{K=1' \;}} =m_{N_{3/2}^{K=1' \;}}=  m_{\Delta_{1/2}^{K=1'}}=  m_{\Delta_{3/2}^{K=1'}}
=  m_{\Delta_{5/2}^{K=1'}}\ , \nonumber \\  
m_{2'} &=&  m_{N_{3/2}^{K=2' \;}} =m_{N_{5/2}^{K=2' \;}}=  m_{\Delta_{1/2}^{K=2'}}=  m_{\Delta_{3/2}^{K=2'}}
=  m_{\Delta_{5/2}^{K=2'}} =  m_{\Delta_{7/2}^{K=2'}} \ , \nonumber \\  
m_{1 \; } &=& m_{N_{1/2}^{K=1 \;}} =m_{N_{3/2}^{K=1 \;}}=  m_{\Delta_{1/2}^{K=1}}=  m_{\Delta_{3/2}^{K=1}}
=  m_{\Delta_{5/2}^{K=1}}  \ , \nonumber \\  
m_{2 \; } &=& m_{N_{3/2}^{K=2 \;}} =m_{N_{5/2}^{K=2 \;}}=  m_{\Delta_{1/2}^{K=2}}=  m_{\Delta_{3/2}^{K=2}}
=  m_{\Delta_{5/2}^{K=2}} =  m_{\Delta_{7/2}^{K=2}} \ , \nonumber \\  
m_{3 \; } &=& m_{N_{5/2}^{K=3 \;}} =m_{N_{7/2}^{K=3 \;}}=  m_{\Delta_{3/2}^{K=3}}=  m_{\Delta_{5/2}^{K=3}}
=  m_{\Delta_{7/2}^{K=3}} =  m_{\Delta_{9/2}^{K=3}}  \ .
\end{eqnarray}
The ``tower masses" $m_K$ correspond to the tower states listed in Eq.~(\ref{Kstates}) and the
degeneracy in the spectrum reflects the $SU(4)_c$ symmetry present in the large $N_c$
limit.
Notice that there are two towers with labels $K=1$ and $K=2$, 
their masses are unrelated by the $SU(4)_c$ symmetry.
The explicit expressions we obtain for the tower masses $m_K$ in terms of the coefficients 
$c_1^{\mathbf{R}},  c_2^{\mathbf{R},\mathbf{R'}}$ and $c_3^{\mathbf{R},\mathbf{R'}}$ result 
from the diagonalization of the mass matrices given in Sec.~\ref{sec:massmat} and can be 
written in compact form as
\begin{eqnarray}
m_{0 \; } &=& N_c \, c_1^{S'_0}\ , \nonumber \\  
m_{1'}    &=& {\overline m}_{11'} + \delta_{11'}\ , \nonumber  \\
m_{2'}    &=& {\overline m}_{22'} + \delta_{22'}\ ,  \\
m_{1 \; } &=&  {\overline m}_{11'} - \delta_{11'}\ , \nonumber  \\
m_{2 \; } &=&  {\overline m}_{22'} - \delta_{22'}\ , \nonumber  \\
m_{3 \; } &=& N_c \, c_1^{MS_2}+c_2^{MS_2}-\frac{2}{7}c_3^{MS_2} \ , \nonumber 
\end{eqnarray}
where
\begin{eqnarray}
\overline m_{11'} &=& \frac{1}{2} (c_1^{MS_0}+ c_1^{MS_2}) N_c -\frac{3}{4}c_2^{MS_2}-\frac{1}{2}c_3^{MS_2} \ , \\ 
\overline m_{22'} &=& \frac{1}{2} (c_1^{S_2}+ c_1^{MS_2}) N_c -\frac{1}{4}c_2^{MS_2}+\frac{1}{2}c_3^{MS_2}  \ , \\ 
\delta_{11'} &=& \sqrt{ \left[\frac{1}{2} (c_1^{MS_0}- c_1^{MS_2}) N_c +\frac{3}{4}c_2^{MS_2}+\frac{1}{2}c_3^{MS_2} \right]^2 
+ 2 \left(c_3^{MS_0MS_2}\right)^2} \ , \\
\delta_{22'} &=& \sqrt{ \left[\frac{1}{2} (c_1^{S_2}- c_1^{MS_2}) N_c +\frac{1}{4}c_2^{MS_2}-\frac{1}{2}c_3^{MS_2} \right]^2 
+ 2 \left(c_2^{S_2MS_2}\right)^2} \ .
\end{eqnarray}
Given the complexity of the mass matrices from where we started, 
these expressions are surprisingly simple. It is possible to understand this by looking at the 
general structure of the corresponding eigenstates shown in the previous Section: 
only some subset of spin-flavor 
states are mixed among each other in the large $N_c$ limit, namely those with same $K$ assignment. 
As we will show next by working out an explicit example, all our results can be understood as a two level $K,K'$ mixing. 
To make this manifest it is useful to write the tower masses $m_K$ we obtained in terms of the mass eigenvalues $\mathring m_K$
that we would have in the absence of configuration mixing 
\begin{eqnarray}
\mathring m_{K=0 \; } &=& N_c \, c_1^{S'_0} \ , \nonumber \\  
\mathring m_{K=1'} &=& N_c \, c_1^{MS_0}	\ , \nonumber \\  
\mathring m_{K=2'} &=& N_c \, c_1^{S_2} \ , \\  
\mathring m_{K=1 \; } &=& N_c \, c_1^{MS_2}-\frac{3}{2}c_2^{MS_2}-c_3^{MS_2} \ ,\nonumber \\  
\mathring m_{K=2 \; } &=& N_c \, c_1^{MS_2}-\frac{1}{2}c_2^{MS_2}+c_3^{MS_2} \ , \nonumber \\  
\mathring m_{K=3 \; } &=& N_c \, c_1^{MS_2}+c_2^{MS_2}-\frac{2}{7}c_3^{MS_2} \ . \nonumber 
\end{eqnarray}
We see that $m_K = \mathring m_K$ for $K=0,3$ and for the $K=1,2$ states we have
\begin{eqnarray}
m_{K',K} &=& \frac{\mathring m_{K}+ \mathring m_{K'}}{2} \pm \sqrt{ \left(\frac{\mathring m_{K'}-\mathring m_{K}}{2}\right)^2 + \left({\mu_K}\right)^2} \ , 
\end{eqnarray}
where $\mu_{K=1}=-\sqrt{2} c_3^{MS_0MS_2}$ and $\mu_{K=2}=-\sqrt{2} c_2^{S_2MS_2}$ are the matrix elements 
that mix the two $K$ states. 
For  $(m_{K}-m_{K'}) \sim {\cal O}(N_c^0)$,  
i.e. $c_1^{MS_0}-c_1^{MS_2} \sim {\cal O}(1/N_c) $ and  $c_1^{MS_2}-c_1^{S_2} \sim {\cal O}(1/N_c) $, 
the mixing is strong and the energy levels get ${\cal O}(N_c^0) $ corrections due to configuration 
mixing.

This mixing pattern, which is not obvious at all when starting from the mass matrices written in the 
spin-flavor basis, can be made manifest by a change of basis. The $I=3/2, J=3/2$ states constitute a 
good example to see this, as they have the  
largest mass matrix, of dimension six and given by Eq.~(\ref{example32}), 
where all $K$ states appear as eigenstates.
To find the change of basis we need, we first compute 
the eigenstates in the absence of mixing by setting $c_{2,3}^{\mathbf{R},\mathbf{R'}}=0$ in Eq.~(\ref{example32}).
In our matrix notation they are given by the row vectors of the following expression 
\begin{eqnarray}
\mathring S &=&
\left(
\begin{array}{cccccc}
1 & 	0 & 		0 & 		0 & 			0 & 			0 \\
0 &	1  &	0 &		0 & 0 & 0	 \\
0 &	0 &	 1 &		0 & 0 & 0	\\
0 & 0 &	0 &		\frac{1}{2\sqrt{5}} & -\frac{2 \sqrt{2} }{5}&		-\frac{3 \sqrt{7} }{10} \\
0 &	0 &		0  &	\frac{1}{2}&	-\sqrt{\frac{2}{5}}&	\frac{1}{2}\sqrt{\frac{7}{5}}\\
0 &	0 &		0 &		-\sqrt{\frac{7}{10}} &		-\frac{\sqrt{7}}{5} &\frac{1}{5 \sqrt{2} }	 \\
\end{array}
\right) \ .
\end{eqnarray}
This provides us the change of basis matrix $\mathring S$. We obtain the large $N_c $ mass matrix $ \tilde M_{{\Delta_{3/2}}} $ 
in the $K$ basis 
\{
$K=0$,  
$K=1'$,  
$K=2'$,  
$K=1$,  
$K=2$,  
$K=3$  
\}
as follows 
\begin{eqnarray}
\tilde M_{{\Delta_{3/2}}} = \mathring S  M_{{\Delta_{3/2}}} \mathring  S^{-1} &=& 
\left(
\begin{array}{cccccc}
\mathring m_0	 &   &   &   &   &   \\
  & \mathring m_{1'} &   & \mu_1 &   &    \\
  &   &  \mathring m_{2'}	 &    & \mu_2 &   \\
  & \mu_1  &   &  \mathring m_{1} &   &   \\
  &   & \mu_2 &   &  \mathring m_{2} &   \\
  &   &   &   &   &  \mathring m_{3} \\
\end{array}
\right) \ , 
\label{massmatK}
\end{eqnarray}
where we only show matrix elements that are non-zero.

The eigenstates in this $K$ basis are given now by the rows of the $T$ matrix below and 
take the simple form
\begin{eqnarray}
T  &=& 
\left(
\begin{array}{cccccc}
1	 &   &   &   &   &   \\
  & 1 &   & \eta_{MS_0} &   &    \\
  &   &  1	 &    & \eta_{S_2} &   \\
  & -\eta_{MS_0}  &   &  1 &   &   \\
  &   & -\eta_{S_2} &   &  1 &   \\
  &   &   &   &   &  1 \\
\end{array}
\right) \ .
\label{statematK}
\end{eqnarray}
Finally, from $S = T \mathring S  $ we recover
as the rows of $S$ the eigenstates in the spin-flavor basis as given by Eq.~(\ref{mD4}). 
\begin{eqnarray}
|\Delta_{3/2}^{K=0} \rangle &=& 
| {^4}\Delta_{3/2}^{S'_0}  \rangle \ , \\
| \Delta_{3/2}^{K=1'}  \rangle &=& 
| {^4}\Delta_{3/2}^{MS_0}  \rangle
+ \eta_{MS_0} \left( 
\frac{1}{2\sqrt{5}} | {^2}\Delta_{3/2}^{MS_2} \rangle
-\frac{2\sqrt{2}}{5} | {^4}\Delta_{3/2}^{MS_2}  \rangle
-\frac{3\sqrt{7}}{10} | {^6}\Delta_{3/2}^{MS_2} \rangle
\right)  \ , \\
| \Delta_{3/2}^{K=2'}  \rangle &=& 
| {^4}\Delta_{3/2}^{S_2}  \rangle
+ \eta_{S_2} \left( 
\frac{1}{2} | {^2}\Delta_{3/2}^{MS_2} \rangle
- \sqrt{\frac{2}{5}} | {^4}\Delta_{3/2}^{MS_2}  \rangle
+ \frac{1}{2}\sqrt{\frac{7}{5}} | {^6}\Delta_{3/2}^{MS_2} \rangle
\right) \ ,  \\
| \Delta_{3/2}^{K=1}  \rangle &=& 
-\eta_{MS_0}| {^4}\Delta_{3/2}^{MS_0}  \rangle
+  
\frac{1}{2\sqrt{5}}  | {^2}\Delta_{3/2}^{MS_2} \rangle
-\frac{2\sqrt{2}}{5} | {^4}\Delta_{3/2}^{MS_2}  \rangle
-\frac{3\sqrt{7}}{10} | {^6}\Delta_{3/2}^{MS_2} \rangle  \ , \\
| \Delta_{3/2}^{K=2}  \rangle &=& 
-\eta_{S_2}| {^4}\Delta_{3/2}^{S_2}  \rangle
+  
\frac{1}{2} | {^2}\Delta_{3/2}^{MS_2} \rangle
-\sqrt{\frac{2}{5}} | {^4}\Delta_{3/2}^{MS_2}  \rangle
+ \frac{1}{2}\sqrt{\frac{7}{5}} | {^6}\Delta_{3/2}^{MS_2} \rangle  \ , \\
| \Delta_{3/2}^{K=3}  \rangle &=& 
-\sqrt{\frac{7}{10}}  | {^2}\Delta_{3/2}^{MS_2} \rangle
-\frac{\sqrt{7}}{5} | {^4}\Delta_{3/2}^{MS_2}  \rangle
+ \frac{1}{5\sqrt{2}} | {^6}\Delta_{3/2}^{MS_2} \rangle  \ .
\end{eqnarray}
The explicit expressions for $ \eta_{MS_0}, \eta_{S_2} $ 
that relate them to the mixing matrix elements $\mu_1, \mu_2$ are
\begin{eqnarray}
\eta_{MS_0}   &=& \frac{2 \mu_1}{\mathring m_{1'}-\mathring m_{1} + 
\sqrt{\left(\mathring m_{1'}-\mathring m_{1}\right)^2 + 4 \left(\mu_1\right)^2 } } \ , \\[2mm] 
\eta_{S_2}   &=& \frac{2 \mu_2}{\mathring m_{2'}-\mathring m_{2} + 
\sqrt{\left(\mathring m_{2'}-\mathring m_{2}\right)^2 + 4 \left(\mu_2\right)^2 } } \ .
\end{eqnarray}

As it is clear from Eqs.~(\ref{massmatK},\ref{statematK}), 
only the two $K=1$ and $K=2$ towers mix through the  mixing matrix elements
$\mu_{K=1}$ and $\mu_{K=2}$, respectively.
This explains why in the large $N_c$ limit all the mixing that can occur among the spin-flavor states
of Eq.~(\ref{states}) can be expressed 
in terms of only two parameters, and it confirms that the effective mass operator given by Eq.~(\ref{massop}) 
correctly accounts for the symmetry structure expected in 
large $N_c$ QCD.

\section{Conclusions}
\label{sec:conclusions}
We have extended the large $N_c$ analysis of excited baryons to include
configuration mixing.  In particular, in this paper we studied the
configuration mixing of the symmetric and mixed symmetric spin-flavor irreps that belong to
the $N=2$ band.  Rather than the $SU(2F)$ spin-flavor symmetry of the quark
model, in the large $N_c$ limit we have a contracted symmetry $SU(2F)_c$, which
gives rise to numerous mass degeneracies and also fixes the mixing pattern
among the states. Degenerate states fill $SU(2F)_c$ irreps (``towers"), which are labeled by $K$. 
They contain an infinite number of states with
increasing spin and isospin. Here we restricted ourselves to two flavors and
to the low spin and isospin states that are identified with the physical states at
$N_c=3$.

We found by an explicit calculation that, in contrast to the complex mixing
pattern of spin-flavor irreps in the quark model due to hyperfine interactions
(see e.g. Ref.~\cite{Isgur:1978wd}), in
the large $N_c$ limit the mixing pattern is governed by the symmetry: 
only states carrying the same $K$ label mix.  For the
states considered, this mixing can be described by just two parameters related to
the mixing of the mixed symmetric states with $L=0$ and $L=2$ ($MS_0$ and
$MS_2$) and the mixing of the symmetric and mixed symmetric states with $L=2$ ($S_2$ and
$MS_2$).

We performed the calculation of the mass spectrum by slightly extending the construction 
of Ref.~\cite{Carlson:1998vx} and using a common 
mass operator for all states, Eq.~(\ref{massop}), 
expressed in the quark operator basis.
We verified explicitly that 
this simple leading order effective mass operator correctly 
describes the configuration mixing pattern expected from the symmetry present in the large 
$N_c$ limit. We also checked explicitly that
the inclusion of configuration mixing preserves the 
large $N_c$ tower 
structure in the spectrum of positive parity excited baryons: The 11 $N$-states 
and 19 $\Delta$-states are the lowest isospin members of six degenerate towers of 
large $N_c$ states labeled by
$K=0,1,1',2,2',3$.   
The matrix elements presented in App.~\ref{app:B} show that ghost states that only
exist for $N_c>3$ decouple from the physical states. This decoupling is a general feature of 
large $N_c$ calculations \cite{Pirjol:2003ye}, that was also pointed out in the 
 meson-baryon scattering picture  \cite{Cohen:2006en}.
Another important point to note is that only the presence of core operators makes 
the mixing of symmetric and mixed-symmetric states possible. A mass operator 
constructed solely in terms of symmetric $SU(4)$ generators $S^i,T^a,G^{ia}$ would not mix spin-flavor states
in different irreps of the permutation group.

All these are non-trivial checks of the correctness of the core-excited quark
picture and the operator construction in terms of $SU(2F)$ generators that
started the program of applying the large $N_c$ expansion to the study of
excited baryons \cite{Goity:1996hk} \cite{Carlson:1998vx}.  Most importantly,
the leading order calculation presented here provides the first step towards a
systematic inclusion of configuration mixing effects at subleading order in the
large $N_c$ analysis of the phenomenology of excited baryons. The predicted
large $N_c$ spectrum and the configuration mixing pattern could also be checked
in the future by lattice calculations, providing a useful guide for the ongoing
explorations of the baryon spectrum at arbitrary $N_c$ values
\cite{DeGrand:2016pur}.

\section*{Acknowledgements}
C.S. and C.W. thank the kind hospitality of the Institut f\"ur Theoretische
Physik II, University of Bochum, were this work was initiated.  We both thank 
Rodolfo Sassot for his support and encouragement.

\pagebreak
\appendix
\section{Calculation of matrix elements}
\label{app:A}
After replacing the orbital angular momentum operator $\ell$ by a generic spatial operator $\xi$, 
we obtain the generalization of equation (A7) of Ref.~\cite{Carlson:1998vx}:
\begin{eqnarray}
\label{o2xi}
 \langle \xi \cdot s \rangle &=& \,\delta_{J'J}\delta_{M'M}\delta_{I'I}\delta_{I_3'I_3}
 (-1)^{L'+1/2+S'-S} \sqrt{\frac{3}{2}} \sqrt{(2S'+1)(2S+1)}
 \langle L'||\xi||L\rangle \nonumber\\
&&\!\!\!\!\!\!\!\!\!\!\!\!\!\!\!\!\!\!\!\!\! \times\sum_{L_s=L\pm 1/2}(-1)^{L_s}(2L_s+1)
 \left\{\begin{array}{ccc}
        L_s & \frac{1}{2} & L' \\
	1   & L   & \frac{1}{2}
      \end{array}\right\}
 \sum_{\eta=\pm1}c^{\mathbf{R}}_{\rho\,\eta}c^{\mathbf{R'}}_{\rho'\eta} 
      \left\{\begin{array}{ccc}
        I_c   & \frac{1}{2} & S' \\
	L & J   & L_s
      \end{array}\right\} 
      \left\{\begin{array}{ccc}
        I_c & \frac{1}{2} & S   \\
	L   &   J         & L_s
      \end{array}\right\}, 
\end{eqnarray}
where $\mathbf{R},\mathbf{R'}$ denote a symmetric ($SYM$) or mixed symmetric ($MS$) irrep and the 
coefficients $c^{\mathbf{R}}_{\rho\,\eta}$ with $S=I+\rho$ and $S_c=I_c=I+\eta/2$ are given by
\begin{eqnarray}
c_{\pm\pm}^{MS} &=& 1 \ , \\
c_{\pm\pm}^{SYM} = c_{\pm\mp}^{SYM} = c_{\pm\mp}^{MS} &=& 0 \ , \\
 c_{0-}^{SYM} = c_{0+}^{MS} &=& \sqrt{\frac{S\left(N_c+2\left(S+1\right)\right)}{N_c\left(2S+1\right)}} \ , \\
 c_{0+}^{SYM} = -c_{0-}^{MS}&=& \sqrt{\frac{\left(S+1\right)\left(N_c-2S\right)}{N_c\left(2S+1\right)}} \ .
\end{eqnarray}
In all other Sections we use $S$ to label the symmetric irrep $SYM$. Here, for the sake of clarity, we 
use the $SYM$ label to distinguish it from the $S$ we also use to denote the total spin of the quarks. 
The rank two tensor operator in the effective mass operator accounts for the mixing of $L=0$ and 
$L=2$ states. It generalizes Eq.~(A9) in Ref.~\cite{Carlson:1998vx} to 
\begin{eqnarray}
\label{o3xi}
 \langle \xi^{(2)} \cdot g \cdot G_c \rangle = &\, \delta_{J'J} \delta_{M'M} \delta_{I'I} \delta_{I_3'I_3} 
(-1)^{J-2I+L+S} \frac{3}{8}\sqrt{5} \sqrt{(2S'+1)(2S+1)} \langle L'||\xi^{(2)}||L\rangle \nonumber \\
&\times
 \left\{\begin{array}{ccc}
	2 & L  & L' \\
	J & S' & S
      \end{array}\right\} 
{\mathlarger{\sum}_{\eta'\eta=\pm1}}c^{\mathbf{R'}}_{\rho'\eta'}c^{\mathbf{R}}_{\rho\,\eta} (-1)^{(1+\eta')/2}\sqrt{(2I_c'+1)(2I_c+1)}
\nonumber \\
&\times
\sqrt{(N_c+1)^2-\left(\frac{\eta'-\eta}{2}\right)^2(2I+1)^2}
 \left\{\begin{array}{ccc}
	\frac{1}{2} & 1  & \frac{1}{2} \\
	I_c'& I  & I_c
      \end{array}\right\}
 \left\{\begin{array}{ccc}
	I_c' & I_c  & 1 \\
	S'   & S    & 2 \\
	\frac{1}{2}  & \frac{1}{2}  & 1
      \end{array}\right\} \ .
\end{eqnarray}
The reduced matrix elements $\langle L'||\xi||L\rangle$ and $\langle L'||\xi^{(2)}||L\rangle$  
are unknown and can be absorbed in the operator coefficients of the $1/N_c$ expansion. 

\section{Explicit matrix elements for arbitrary $N_c$}
\label{app:B}

We list in this Appendix all the explicit matrix elements for the operators 
${\cal O}_1 = N_c \openone$, 
${\cal O}_2 = \xi \cdot s $ and 
${\cal O}_3 = \frac{1}{N_c} \xi^{(2)} \cdot g \cdot G_c$, 
for finite $N_c$. We defined  
$\xi_{122}=\frac{1}{\sqrt{30}}\langle 2||\xi||2\rangle$, 
$\xi_{222}=\frac{1}{\sqrt{105}}\langle 2||\xi^{(2)}||2\rangle$
 and $\xi_{202}=\frac{1}{16\sqrt{2}}\langle 0||\xi^{(2)}||2\rangle$, which 
contain the reduced matrix elements of the generic $\xi$ operator.
Note that the ghost $\Delta$ states decouple from the physical states 
through $N_c-3$ factors, see also \cite{Cohen:2006en} for a 
related discussion. 
\begin{table}[h!]
\medskip
\begin{tabular}{cccc}
\hline
\hline
& ${\cal O}_1$ & ${\cal O}_2$ & ${\cal O}_3$\\
\hline
$^2N^{S'_0}_{1/2}$ 			& $N_c$	& $0$						& $0$	\\
$^2N^{MS_0}_{1/2}$ 			& $N_c$	& $0$			 			& $0$ 	\\
$^4N^{MS_2}_{1/2}$ 			& $N_c$	& $-\frac{3}{2} \xi_{122}$			& $-\frac{7}{16N_c}(N_c+1)\xi_{222}$	\\

$^2N^{S'_0}_{1/2}-{}^2N^{MS_0}_{1/2}$	& $0$	& $0$						& $0$	\\
$^2N^{S'_0}_{1/2}-{}^4N^{MS_2}_{1/2}$	& $0$	& $0$						& $-\frac{1}{N_c}\sqrt{\frac{2(N_c-1)}{N_c}}\xi_{202}$	\\

$^2N^{MS_0}_{1/2}-{}^4N^{MS_2}_{1/2}$	& $0$	& $0$						& $\frac{1}{N_c}(2N_c-1)\sqrt{\frac{2(N_c+3)}{3N_c}}\xi_{202}$	\\
\hline
$^4N^{MS_0}_{3/2}$ 			& $N_c$	& 	$0$					& $0$	\\
$^2N^{S_2}_{3/2}$ 			& $N_c$	& $-\frac{3}{2N_c}\xi_{122}$ 			& $0$ 	\\
$^2N^{MS_2}_{3/2}$ 			& $N_c$	& $-\frac{1}{2N_c} (2N_c-3)\xi_{122}$	& $0$	\\
$^4N^{MS_2}_{3/2}$ 			& $N_c$	& $-\xi_{122}$ 					& $0$	\\

$^4N^{MS_0}_{3/2}-{}^2N^{S_2}_{3/2}$	& $0$	& $0$						& $\frac{1}{N_c}\sqrt{\frac{N_c-1}{N_c}}\xi_{202}$	\\
$^4N^{MS_0}_{3/2}-{}^2N^{MS_2}_{3/2}$	& $0$	& $0$						& $-\frac{1}{\sqrt{3}N_c}(2N_c-1)\sqrt{\frac{N_c+3}{N_c}}\xi_{202}$	\\
$^4N^{MS_0}_{3/2}-{}^4N^{MS_2}_{3/2}$	& $0$	& $0$						& $-\frac{2}{\sqrt{3}N_c}(N_c+1)\xi_{202}$	\\

$^2N^{S_2}_{3/2}-{}^2N^{MS_2}_{3/2}$	& $0$	& $\frac{1}{2N_c}\sqrt{3(N_c+3)(N_c-1)}\xi_{122}$	& $0$	\\
$^2N^{S_2}_{3/2}-{}^4N^{MS_2}_{3/2}$	& $0$	& $-\frac{1}{2}\sqrt{\frac{3(N_c-1)}{N_c}}\xi_{122}$	& $\frac{7}{32N_c}\sqrt{\frac{3(N_c-1)}{N_c}}\xi_{222}$	\\

$^2N^{MS_2}_{3/2}-{}^4N^{MS_2}_{3/2}$	& $0$	& $-\frac{1}{2}\sqrt{\frac{N_c+3}{N_c}}\xi_{122}$	& $-\frac{7}{32N_c}(2N_c-1)\sqrt{\frac{N_c+3}{N_c}}\xi_{222}$\\\hline
$^2N^{S_2}_{5/2}$ 			& $N_c$	& $\frac{1}{N_c}\xi_{122}$			& $0$	\\
$^2N^{MS_2}_{5/2}$ 			& $N_c$	& $\frac{1}{3 N_c}(2 N_c-3) \xi_{122}$		& $0$ 	\\
$^4N^{MS_2}_{5/2}$ 			& $N_c$	& $-\frac{1}{6}\xi_{122}$			& $\frac{5}{16 N_c}(N_c+1) \xi_{222}$	\\

$^2N^{S_2}_{5/2}-{}^2N^{MS_2}_{5/2}$	& $0$	& $-\frac{1}{3N_c}\sqrt{(N_c+3)(N_c-1)} \xi_{122}$		& $0$	\\
$^2N^{S_2}_{5/2}-{}^4N^{MS_2}_{5/2}$	& $0$	& $-\sqrt{\frac{7}{6}} \sqrt{\frac{N_c-1}{N_c}} \xi_{122}$	& $-\frac{1}{16N_c} \sqrt{\frac{21}{2}} \sqrt{\frac{N_c-1}{N_c}} \xi_{222}$	\\

$^2N^{MS_2}_{5/2}-{}^4N^{MS_2}_{5/2}$	& $0$	& $-\frac{1}{3} \sqrt{\frac{7}{2}} \sqrt{\frac{N_c+3}{N_c}} \xi_{122}$	& $\sqrt{\frac{7}{2}}\frac{ (2 N_c-1)}{16 N_c} \sqrt{\frac{N_c+3}{N_c}} \xi_{222}$\\\hline
$^4N^{MS_2}_{7/2}$ 			& $N_c$	& $\xi_{122}$			& $-\frac{1}{8 N_c} (N_c+1)\xi_{222}$	\\
\hline\hline
\end{tabular}
\caption{Matrix elements for $I=1/2$ states at finite $N_c$.}\label{table12}
\end{table}

\begin{table}
\medskip
\begin{tabular}{cccc}
\hline
\hline
& ${\cal O}_1$ & ${\cal O}_2$ & ${\cal O}_3$\\
\hline
$^2\Delta^{MS_0}_{1/2}$			& $N_c$	& $0$						& $0$	\\
$^4\Delta^{S_2}_{1/2}$ 			& $N_c$	& $-\frac{9}{2N_c}\xi_{122}$		& $\frac{21}{8N_c^2}\xi_{222}$ 	\\
$^4\Delta^{MS_2}_{1/2}$			& $N_c$	& $\frac{3}{10 N_c}(15-2 N_c)\xi_{122}$	& $\frac{7}{40 N_c^2}(2 N_c^2+2 N_c-15) \xi_{222}$\\
$^6\Delta^{MS_2}_{1/2}$			& $N_c$	& $-\frac{7}{5}\xi_{122}$			& $-\frac{7}{20 N_c}(N_c+1) \xi_{222}$	\\

$^2\Delta^{MS_0}_{1/2}-{}^4\Delta^{S_2}_{1/2}$	& $0$	& $0$						& $-\frac{1}{N_c} \sqrt{\frac{N_c+5}{N_c}} \xi_{202}$	\\
$^2\Delta^{MS_0}_{1/2}-{}^4\Delta^{MS_2}_{1/2}$	& $0$	& $0$						& $\frac{1}{\sqrt{15} N_c} (2N_c+5)\sqrt{\frac{N_c-3}{N_c}} \xi_{202}$	\\
$^2\Delta^{MS_0}_{1/2}-{}^6\Delta^{MS_2}_{1/2}$	& $0$	& $0$						& $\frac{2}{N_c} \sqrt{\frac{3}{5}} \sqrt{(N_c+5)(N_c-3)}\xi_{202}$	\\

$^4\Delta^{S_2}_{1/2}-{}^4\Delta^{MS_2}_{1/2}$	& $0$	& $\frac{3}{2N_c} \sqrt{\frac{3}{5}} \sqrt{(N_c+5)(N_c-3)} \xi_{122}$		& $-\frac{7}{8N_c^2} \sqrt{\frac{3}{5}}\sqrt{(N_c+5)(N_c-3)} \xi_{222}$	\\
$^4\Delta^{S_2}_{1/2}-{}^6\Delta^{MS_2}_{1/2}$	& $0$	& $-\frac{1}{2} \sqrt{\frac{3}{5}} \sqrt{\frac{N_c-3}{N_c}} \xi_{122}$	& $\frac{21}{32N_c} \sqrt{\frac{3}{5}}\sqrt{\frac{N_c-3}{N_c}} \xi_{222}$	\\

$^4\Delta^{MS_2}_{1/2}-{}^6\Delta^{MS_2}_{1/2}$	& $0$	& $-\frac{3}{10} \sqrt{\frac{N_c+5}{N_c}} \xi_{122}$						& $-\frac{21}{160 N_c}(2N_c-3)\sqrt{\frac{N_c+5}{N_c}} \xi_{222}$	\\
\hline\hline
\end{tabular}
\caption{Matrix elements for the $I=3/2$, $J=1/2$ states at finite $N_c$.}\label{table3212}
\end{table}

\begin{table}
\medskip
\begin{tabular}{cccc}
\hline
\hline
& ${\cal O}_1$ & ${\cal O}_2$ & ${\cal O}_3$\\
\hline
$^4\Delta^{S_0}_{3/2}$ 			& $N_c$	& $0$						& $0$	\\
$^4\Delta^{MS_0}_{3/2}$ 		& $N_c$	& $0$			 			& $0$ 	\\
$^4\Delta^{S_2}_{3/2}$ 			& $N_c$	& $-\frac{3}{N_c}\xi_{122}$			& $0$	\\
$^2\Delta^{MS_2}_{3/2}$ 		& $N_c$	& $\frac{1}{2}\xi_{122}$				& $0$	\\
$^4\Delta^{MS_2}_{3/2}$ 		& $N_c$	& $-\frac{1}{5N_c}(2N_c-15)\xi_{122}$		& $0$	\\
$^6\Delta^{MS_2}_{3/2}$ 		& $N_c$	& $-\frac{11}{10}\xi_{122}$			& $-\frac{1}{8N_c}(N_c+1)\xi_{222}$\\

$^4\Delta^{S'_0}_{3/2}-{}^4\Delta^{MS_0}_{3/2}$ 	& $0$	& $0$					& $0$	\\
$^4\Delta^{S'_0}_{3/2}-{}^4\Delta^{S_2}_{3/2}$ 	& $0$	& $0$					& $\frac{4\sqrt{3}}{N_c^2}\xi_{202}$\\
$^4\Delta^{S'_0}_{3/2}-{}^2\Delta^{MS_2}_{3/2}$ 	& $0$	& $0$					& $ \frac{1}{\sqrt{2}N_c	}\sqrt{\frac{N_c+5}{N_c}} \xi_{202}$\\
$^4\Delta^{S'_0}_{3/2}-{}^4\Delta^{MS_2}_{3/2}$ 	& $0$	& $0$					& $-\frac{4}{\sqrt{5} N_c^2}\sqrt{(N_c+5)(N_c-3)} \xi_{202}$\\
$^4\Delta^{S'_0}_{3/2}-{}^6\Delta^{MS_2}_{3/2}$ 	& $0$	& $0$					& $-\frac{3}{N_c} \sqrt{\frac{7}{10}}\sqrt{\frac{N_c-3}{N_c}}\xi_{202}$\\

$^4\Delta^{MS_0}_{3/2}-{}^4\Delta^{S_2}_{3/2}$ 		& $0$	& $0$	& $-\frac{4}{\sqrt{5} N_c^2} \sqrt{(N_c+5)(N_c-3)} \xi_{202}$\\
$^4\Delta^{MS_0}_{3/2}-{}^2\Delta^{MS_2}_{3/2}$ 	& $0$	& $0$	& $-\frac{1}{\sqrt{30}N_c} (2 N_c+5) \sqrt{\frac{N_c-3}{N_c}} \xi_{202}$\\
$^4\Delta^{MS_0}_{3/2}-{}^4\Delta^{MS_2}_{3/2}$ 	& $0$	& $0$	& $\frac{4}{5 \sqrt{3} N_c^2} \left(2 N_c^2+2 N_c-15\right)\xi_{202}$\\
$^4\Delta^{MS_0}_{3/2}-{}^6\Delta^{MS_2}_{3/2}$ 	& $0$	& $0$	& $\frac{1}{5 N_c}\sqrt{\frac{21}{2}}(2 N_c-3) \sqrt{\frac{N_c+5}{N_c}} \xi_{202}$\\

$^4\Delta^{S_2}_{3/2}-{}^2\Delta^{MS_2}_{3/2}$ 		& $0$	& $-\frac{1}{2} \sqrt{\frac{3}{2}} \sqrt{\frac{N_c+5}{N_c}} \xi_{122}$	& $ \frac{7}{32N_c} \sqrt{\frac{3}{2}} \sqrt{\frac{N_c+5}{N_c}} \xi_{222}$\\
$^4\Delta^{S_2}_{3/2}-{}^4\Delta^{MS_2}_{3/2}$ 		& $0$	& $\frac{1}{N_c}\sqrt{\frac{3}{5}} \sqrt{(N_c+5)(N_c-3)} \xi_{122}$		& $0$\\
$^4\Delta^{S_2}_{3/2}-{}^6\Delta^{MS_2}_{3/2}$ 		& $0$	& $-\frac{1}{2} \sqrt{\frac{21}{10}} \sqrt{\frac{N_c-3}{N_c}} \xi_{122}$	& $\frac{3}{32 N_c}\sqrt{\frac{105}{2}} \sqrt{\frac{N_c-3}{N_c}} \xi_{222}$\\

$^2\Delta^{MS_2}_{3/2}-{}^4\Delta^{MS_2}_{3/2}$ 	& $0$	& $\frac{1}{2} \sqrt{\frac{5}{2}} \sqrt{\frac{N_c-3}{N_c}} \xi_{122}$	& $-\frac{7}{32\sqrt{10}N_c} (2 N_c+5)  \sqrt{\frac{N_c-3}{N_c}}\xi_{222}$\\
$^2\Delta^{MS_2}_{3/2}-{}^6\Delta^{MS_2}_{3/2}$ 	& $0$	& $0$								& $\frac{3}{16N_c}\sqrt{\frac{7}{5}}\sqrt{(N_c+5)(N_c-3)}\xi_{222}$\\

$^4\Delta^{MS_2}_{3/2}-{}^6\Delta^{MS_2}_{3/2}$ 	& $0$	& $-\frac{3}{10} \sqrt{\frac{7}{2}} \sqrt{\frac{N_c+5}{N_c}} \xi_{122}$	& $-\frac{3}{32N_c}\sqrt{\frac{7}{2}}(2 N_c-3)\sqrt{\frac{N_c+5}{N_c}} \xi_{222}$ \\
\hline\hline
\end{tabular}
\caption{Matrix elements for the $I=3/2$, $J=3/2$ states at finite $N_c$.}\label{table3232}
\end{table}

\begin{table}
\medskip
\begin{tabular}{cccc}
\hline
\hline
& ${\cal O}_1$ & ${\cal O}_2$ & ${\cal O}_3$\\
\hline
$^6\Delta^{MS_0}_{5/2}$			& $N_c$	& $0$					& $0$	\\
$^4\Delta^{S_2}_{5/2}$ 			& $N_c$	& $-\frac{1}{2N_c}\xi_{122}$		& $-\frac{15}{8 N_c^2} \xi_{222}$ 	\\
$^2\Delta^{MS_2}_{5/2}$			& $N_c$	& $-\frac{1}{3}\xi_{122}$			& $0$	\\
$^4\Delta^{MS_2}_{5/2}$			& $N_c$	& $-\frac{1}{30 N_c}(2 N_c-15) \xi_{122}$	& $-\frac{1}{8 N_c^2}(2N_c^2+2N_c-15) \xi_{222}$	\\
$^6\Delta^{MS_2}_{5/2}$			& $N_c$	& $-\frac{3}{5} \xi_{122}$		& $\frac{1}{8 N_c}(N_c+1)\xi_{222}$	\\

$^6\Delta^{MS_0}_{5/2}-{}^4\Delta^{S_2}_{5/2}$	& $0$	& $0$			& $\frac{1}{N_c}\sqrt{\frac{21}{5}}\sqrt{\frac{N_c-3}{N_c}}\xi_{202}$	\\
$^6\Delta^{MS_0}_{5/2}-{}^2\Delta^{MS_2}_{5/2}$	& $0$	& $0$			& $\frac{2}{\sqrt{5}N_c}\sqrt{(N_c+5)(N_c-3)} \xi_{202}$	\\
$^6\Delta^{MS_0}_{5/2}-{}^4\Delta^{MS_2}_{5/2}$	& $0$	& $0$			& $-\frac{\sqrt{7}}{5N_c}(2N_c-3)\sqrt{\frac{N_c+5}{N_c}}\xi_{202}$\\
$^6\Delta^{MS_0}_{5/2}-{}^6\Delta^{MS_2}_{5/2}$	& $0$	& $0$			& $-\frac{2}{5 N_c} \sqrt{\frac{14}{3}}(N_c+1) \xi_{202}$	\\

$^4\Delta^{S_2}_{5/2}-{}^2\Delta^{MS_2}_{5/2}$	& $0$	& $-\frac{1}{2} \sqrt{\frac{7}{3}} \sqrt{\frac{N_c+5}{N_c}} \xi_{122}$	& $-\frac{\sqrt{21}}{32N_c}\sqrt{\frac{N_c+5}{N_c}}\xi_{222}$\\
$^4\Delta^{S_2}_{5/2}-{}^4\Delta^{MS_2}_{5/2}$	& $0$	& $\frac{1}{2\sqrt{15}N_c}\sqrt{(N_c+5)(N_c-3)} \xi_{122}$			& $\frac{\sqrt{15}}{8 N_c^2} \sqrt{(N_c+5)(N_c-3)} \xi_{222}$	\\
$^4\Delta^{S_2}_{5/2}-{}^6\Delta^{MS_2}_{5/2}$	& $0$	& $-\frac{3}{\sqrt{10}} \sqrt{\frac{N_c-3}{N_c}} \xi_{122}$			& $\frac{3}{16N_c} \sqrt{\frac{5}{2}} \sqrt{\frac{N_c-3}{N_c}} \xi_{222}$	\\

$^2\Delta^{MS_2}_{5/2}-{}^4\Delta^{MS_2}_{5/2}$	& $0$	& $\frac{\sqrt{35}}{6}\sqrt{\frac{N_c-3}{N_c}} \xi_{122}$	& $\frac{1}{32N_c}\sqrt{\frac{7}{5}}(2 N_c+5)\sqrt{\frac{N_c-3}{N_c}}\xi_{222}$	\\
$^2\Delta^{MS_2}_{5/2}-{}^6\Delta^{MS_2}_{5/2}$	& $0$	& $0$							& $\frac{1}{4N_c}\sqrt{\frac{21}{10}}\sqrt{(N_c+5)(N_c-3)}\xi_{222}$	\\

$^4\Delta^{MS_2}_{5/2}-{}^6\Delta^{MS_2}_{5/2}$	& $0$	& $-\frac{3}{5}\sqrt{\frac{3}{2}} \sqrt{\frac{N_c+5}{N_c}} \xi_{122}$		& $-\frac{1}{16N_c}\sqrt{\frac{3}{2}}(2 N_c-3)\sqrt{\frac{N_c+5}{N_c}} \xi_{222}$	\\
\hline
$^4\Delta^{S_2}_{7/2}$ 			& $N_c$	& $\frac{3}{N_c} \xi_{122}$			& $\frac{3}{4 N_c^2} \xi_{222}$	\\
$^4\Delta^{MS_2}_{7/2}$ 		& $N_c$	& $\frac{1}{5 N_c}(2 N_c-15) \xi_{122}$	& $\frac{1}{20 N_c^2}(2 N_c^2+2N_c-15) \xi_{222}$ 	\\
$^6\Delta^{MS_2}_{7/2}$ 		& $N_c$	& $\frac{1}{10}\xi_{122}$			& $\frac{17}{80N_c}(N_c+1)\xi_{222}$	\\

$^4\Delta^{S_2}_{7/2}-{}^4\Delta^{MS_2}_{7/2}$	& $0$	& $-\frac{1}{N_c}\sqrt{\frac{3}{5}} \sqrt{(N_c+5)(N_c-3)}\xi_{122}$	& $-\frac{1}{4 N_c^2}\sqrt{\frac{3}{5}}\sqrt{(N_c+5)(N_c-3)} \xi_{222}$	\\
$^4\Delta^{S_2}_{7/2}-{}^6\Delta^{MS_2}_{7/2}$	& $0$	& $-\frac{3}{\sqrt{10}} \sqrt{\frac{N_c-3}{N_c}} \xi_{122}$		& $-\frac{27}{16\sqrt{10}N_c} \sqrt{\frac{N_c-3}{N_c}} \xi_{222}$	\\

$^4\Delta^{MS_2}_{7/2}-{}^6\Delta^{MS_2}_{7/2}$	& $0$	& $-\frac{3}{5} \sqrt{\frac{3}{2}} \sqrt{\frac{N_c+5}{N_c}} \xi_{122}$	& $ \frac{9}{80N_c}\sqrt{\frac{3}{2}}(2 N_c-3)\sqrt{\frac{N_c+5}{N_c}} \xi_{222}$	\\
\hline
$^6\Delta^{MS_2}_{9/2}$ 	& $N_c$	& $\xi_{122}$			& $-\frac{1}{8 N_c}(N_c+1) \xi_{222}$	\\
\hline\hline
\end{tabular}
\caption{Matrix elements for the $I=3/2$, $J=5/2, 7/2, 9/2$ states at finite $N_c$.}\label{table3252}
\end{table}


\clearpage

\begin{thebibliography}{99}

\bibitem{Capstick:2000qj} 
  S.~Capstick and W.~Roberts,
  Prog.\ Part.\ Nucl.\ Phys.\  {\bf 45}, S241 (2000)
  [nucl-th/0008028].

\bibitem{Klempt:2009pi} 
  E.~Klempt and J.~M.~Richard,
  Rev.\ Mod.\ Phys.\  {\bf 82}, 1095 (2010)
  [arXiv:0901.2055 [hep-ph]].

\bibitem{Crede:2013kia} 
  V.~Crede and W.~Roberts,
  Rept.\ Prog.\ Phys.\  {\bf 76}, 076301 (2013)
  [arXiv:1302.7299 [nucl-ex]].

\bibitem{Edwards:2011jj} 
  R.~G.~Edwards, J.~J.~Dudek, D.~G.~Richards and S.~J.~Wallace,
  Phys.\ Rev.\ D {\bf 84}, 074508 (2011)
  [arXiv:1104.5152 [hep-ph]].

\bibitem{Edwards:2012fx} 
  R.~G.~Edwards {\it et al.}  [Hadron Spectrum Collaboration],
  Phys.\ Rev.\ D {\bf 87}, no. 5, 054506 (2013)
  [arXiv:1212.5236 [hep-ph]].

\bibitem{'tHooft:1973jz} 
  G.~'t Hooft,
  Nucl.\ Phys.\ B {\bf 72}, 461 (1974).

\bibitem{Witten:1979kh} 
  E.~Witten,
  Nucl.\ Phys.\ B {\bf 160}, 57 (1979).

\bibitem{Lucini:2012gg} 
  B.~Lucini and M.~Panero,
  Phys.\ Rept.\  {\bf 526}, 93 (2013)
  [arXiv:1210.4997 [hep-th]].

\bibitem{Gervais:1983wq} 
  J.~L.~Gervais and B.~Sakita,
  Phys.\ Rev.\ Lett.\  {\bf 52}, 87 (1984).

\bibitem{Gervais:1984rc} 
  J.~L.~Gervais and B.~Sakita,
  Phys.\ Rev.\ D {\bf 30}, 1795 (1984).

\bibitem{Bardakci:1983ev} 
  K.~Bardakci,
  Nucl.\ Phys.\ B {\bf 243}, 197 (1984).

\bibitem{Dashen:1993as} 
  R.~F.~Dashen and A.~V.~Manohar,
  Phys.\ Lett.\ B {\bf 315}, 425 (1993)
  [hep-ph/9307241].

\bibitem{Jenkins:1993zu} 
  E.~E.~Jenkins,
  Phys.\ Lett.\ B {\bf 315}, 441 (1993)
  [hep-ph/9307244].

\bibitem{Dashen:1993jt} 
  R.~F.~Dashen, E.~E.~Jenkins and A.~V.~Manohar,
  Phys.\ Rev.\ D {\bf 49}, 4713 (1994)
  [Phys.\ Rev.\ D {\bf 51}, 2489 (1995)]
  [hep-ph/9310379].

\bibitem{Carone:1993dz} 
  C.~Carone, H.~Georgi and S.~Osofsky,
  Phys.\ Lett.\ B {\bf 322}, 227 (1994)
  [hep-ph/9310365].

\bibitem{Luty:1993fu} 
  M.~A.~Luty and J.~March-Russell,
  Nucl.\ Phys.\ B {\bf 426}, 71 (1994)
  [hep-ph/9310369].


\bibitem{Dashen:1994qi} 
  R.~F.~Dashen, E.~E.~Jenkins and A.~V.~Manohar,
  Phys.\ Rev.\ D {\bf 51}, 3697 (1995)
  [hep-ph/9411234].

\bibitem{Manohar:1998xv} 
  A.~V.~Manohar, in 
``Probing the standard model of particle interactions. Proceedings, Summer School in Theoretical Physics, NATO Advanced Study Institute, 68th session, Les Houches, France, July 28-September 5, 1997. Pt. 1, 2,''
 R.~Gupta, A.~Morel, E.~de Rafael and F.~David.
  Amsterdam, Netherlands: Elsevier (1999) 1642 p.
  [hep-ph/9802419].

\bibitem{Jenkins:1998wy} 
  E.~E.~Jenkins,
  Ann.\ Rev.\ Nucl.\ Part.\ Sci.\  {\bf 48}, 81 (1998)
  [hep-ph/9803349].


\bibitem{Lebed:1998st} 
  R.~F.~Lebed,
  Czech.\ J.\ Phys.\  {\bf 49}, 1273 (1999)
  [nucl-th/9810080].
 
\bibitem{Goity:1996hk} 
  J.~L.~Goity,
  Phys.\ Lett.\ B {\bf 414}, 140 (1997)
  [hep-ph/9612252].

\bibitem{Pirjol:1997bp} 
  D.~Pirjol and T.~M.~Yan,
  Phys.\ Rev.\ D {\bf 57}, 1449 (1998)
  [hep-ph/9707485].

\bibitem{Pirjol:1997sr} 
  D.~Pirjol and T.~M.~Yan,
  Phys.\ Rev.\ D {\bf 57}, 5434 (1998)
  [hep-ph/9711201].

\bibitem{Carlson:1998vx} 
  C.~E.~Carlson, C.~D.~Carone, J.~L.~Goity and R.~F.~Lebed,
  Phys.\ Rev.\ D {\bf 59}, 114008 (1999)
  [hep-ph/9812440].

\bibitem{Schat:2001xr} 
  C.~L.~Schat, J.~L.~Goity and N.~N.~Scoccola,
  Phys.\ Rev.\ Lett.\  {\bf 88}, 102002 (2002)
  [hep-ph/0111082].

\bibitem{Goity:2002pu} 
  J.~L.~Goity, C.~L.~Schat and N.~N.~Scoccola,
  Phys.\ Rev.\ D {\bf 66}, 114014 (2002)
  [hep-ph/0209174].

\bibitem{Matagne:2014lla} 
  N.~Matagne and F.~Stancu,
  Rev.\ Mod.\ Phys.\  {\bf 87}, 211 (2015)
  [arXiv:1406.1791 [hep-ph]].

\bibitem{Jenkins:2009wv} 
  E.~E.~Jenkins, A.~V.~Manohar, J.~W.~Negele and A.~Walker-Loud,
  Phys.\ Rev.\ D {\bf 81}, 014502 (2010)
  [arXiv:0907.0529 [hep-lat]].

\bibitem{Fernando:2014dna} 
  I.~P.~Fernando and J.~L.~Goity,
  Phys.\ Rev.\ D {\bf 91}, no. 3, 036005 (2015)
  [arXiv:1410.1384 [hep-ph]].

\bibitem{Cordon:2014sda} 
  A.~C.~Cord\'on, T.~DeGrand and J.~L.~Goity,
  Phys.\ Rev.\ D {\bf 90}, no. 1, 014505 (2014)
  [arXiv:1404.2301 [hep-ph]].

\bibitem{DeGrand:2016pur} 
  T.~DeGrand and Y.~Liu,
  Phys.\ Rev.\ D {\bf 94}, no. 3, 034506 (2016)
  Erratum: [Phys.\ Rev.\ D {\bf 95}, no. 1, 019902 (2017)]
  [arXiv:1606.01277 [hep-lat]].

\bibitem{Pirjol:2003ye} 
  D.~Pirjol and C.~Schat,
  Phys.\ Rev.\ D {\bf 67}, 096009 (2003)
  [hep-ph/0301187].
 
\bibitem{Cohen:2003tb} 
  T.~D.~Cohen and R.~F.~Lebed,
  Phys.\ Rev.\ Lett.\  {\bf 91}, 012001 (2003)
  [hep-ph/0301167].

\bibitem{Matagne:2011sn} 
  N.~Matagne and F.~Stancu,
  Phys.\ Rev.\ D {\bf 84}, 056013 (2011)
  [arXiv:1106.4992 [hep-ph]].

\bibitem{Goity:2004pw} 
  J.~L.~Goity,
  Phys.\ Atom.\ Nucl.\  {\bf 68}, 624 (2005)
  [Yad.\ Fiz.\  {\bf 68}, 655 (2005)]
  doi:10.1134/1.1903092
  [hep-ph/0405304].

\bibitem{Goity:2005gs} 
  J.~L.~Goity, in   ``Large N(c) QCD 2004. Proceedings, International Workshop, Trento, Italy, July 5-11, 2004,''
J.~L.~Goity, R.~F.~Lebed, A.~Pich, C.~L.~Schat and N.~N.~Scoccola. USA: World Scientific (2005) 307 p.
  [hep-ph/0504121].

\bibitem{Goity:2004ss} 
  J.~L.~Goity, C.~Schat and N.~Scoccola,
  Phys.\ Rev.\ D {\bf 71}, 034016 (2005)
  [hep-ph/0411092].

\bibitem{Cohen:2006en} 
  T.~D.~Cohen and R.~F.~Lebed,
  Phys.\ Rev.\ D {\bf 74}, 036001 (2006)
  [hep-ph/0604175].

\bibitem{Isgur:1978wd} 
  N.~Isgur and G.~Karl,
  Phys.\ Rev.\ D {\bf 19}, 2653 (1979)
  Erratum: [Phys.\ Rev.\ D {\bf 23}, 817 (1981)].

\bibitem{Carone:1994tu} 
  C.~D.~Carone, H.~Georgi, L.~Kaplan and D.~Morin,
  Phys.\ Rev.\ D {\bf 50}, 5793 (1994)
  [hep-ph/9406227].

\bibitem{Collins:1998ny}
  H.~Collins and H.~Georgi,
  Phys.\ Rev.\  D {\bf 59}, 094010 (1999)
  [arXiv:hep-ph/9810392].

\bibitem{Pirjol:2007ed} 
  D.~Pirjol and C.~Schat,
  Phys.\ Rev.\ D {\bf 78}, 034026 (2008)
  [arXiv:0709.0714 [hep-ph]].

\bibitem{Galeta:2009pn} 
  L.~Galeta, D.~Pirjol and C.~Schat,
  Phys.\ Rev.\ D {\bf 80}, 116004 (2009)
  [arXiv:0906.0699 [hep-ph]].

\bibitem{Pirjol:2010th} 
  D.~Pirjol and C.~Schat,
  Phys.\ Rev.\ D {\bf 82}, 114005 (2010)
  [arXiv:1007.0964 [hep-ph]].

\bibitem{Willemyns:2015hgy} 
  C.~Willemyns and C.~Schat,
  Phys.\ Rev.\ D {\bf 93}, no. 3, 034007 (2016)
  [arXiv:1511.02215 [nucl-th]].



\end{thebibliography}
\end{document}